\documentclass[twocolumn,twoside,slac_two]{revtex4}
\usepackage{graphicx}
\usepackage{fancyhdr}
\newcommand{\fermi}{\textit{Fermi}-LAT}

\newcommand{\src}{3C~120}

\def\aap{{\em A\&A}\ }

\def\aj{{\em AJ}\ }

\def\apj{{\em ApJ}\ }
\def\apjl{{\em ApJL}\ }

\def\mnras{{\em MNRAS}\ }
\def\nat{{\em Nature}\ }

\def\pasj{{\em PASJ}\ }
\def\pasp{{\em PASP}\ }

\def\procspie{{\em Proc.~SPIE}\ }

\begin{document}

\title{Fermi-LAT and Multi-wavelength Monitoring of the Broad Line Radio Galaxy 3C~120}

%

%


\author{Y.~T.~Tanaka$^1$, A.~Doi$^2$, Y.~Inoue$^2$, C.~C.~Cheung$^3$, L.~Stawarz$^{2, 4}$, Y.~Fukazawa$^5$, M.~A.~Gurwell$^6$, M.~Tahara$^7$, J.~Kataoka$^7$, and R.~Itoh$^5$, on behalf of the Fermi-LAT collaboration}

\affiliation{$^1$Hiroshima Astrophysical Science Center, Hiroshima University, Higashi-Hiroshima, 739-8526, Japan}
\affiliation{$^2$ISAS/JAXA, Sagamihara, Kanagawa, 252-5210, Japan}
\affiliation{$^3$Space Science Division, Naval Research Laboratory, Washington, DC 20375-5352, USA}
\affiliation{$^4$Astronomical Observatory, Jagiellonian University, ul. Orla 171, 30-244 Krak\'ow, Poland}
\affiliation{$^5$Department of Physical Sciences, Hiroshima University, Higashi-Hiroshima, Hiroshima 739-8526, Japan}
\affiliation{$^6$Harvard-Smithsonian Center for Astrophysics, Cambridge, MA 02138, USA}
\affiliation{$^7$Research Institute for Science and Engineering, Waseda University, Tokyo 169-8555, Japan}

\begin{abstract}
We present six-year multi-wavelength monitoring result for broad-line radio galaxy 3C~120. The source was sporadically detected by \fermi\ and after the MeV/GeV $\gamma$-ray detection the 43~GHz radio core brightened and a knot ejected from an unresolved core, implying that the radio-gamma phenomena are physically connected. We show that the $\gamma$-ray emission region is located at sub-pc distance from the central black hole, and MeV/GeV $\gamma$-ray emission mechanism is inverse-Compton scattering of synchrotron photons. We also discuss future perspective revealed by next-generation X-ray satellite {\it Astro-H}.
\end{abstract}

\maketitle

\thispagestyle{fancy}


\section{Broad Line Radio Galaxy 3C~120}
\src, known as broad-line radio galaxy (BLRG) at $z=0.033$, is classified as Fanaroff-Riley Class I (FR I) radio galaxy based on the radio morphology. Since the viewing angle of the jet is not as small as blazars (whose jets are directed toward the Earth), both thermal disk and non-thermal jet components are present in the broad band spectrum. In this regard, \src\ and also 3C~111 are ideal objects to investigate the disk-jet connection. For example, from long-term Very Long Baseline Array (VLBA) and X-ray monitorings, X-ray dimmings are found to be followed by radio knot ejections \cite{Marscher02, Chatterjee09, Chatterjee11}. This phenomenon is understood by a scenario that the disk materials suddenly fall onto the central black hole (BH) and then they are ejected as a jet \cite[e.g.,][]{Marscher02}.

It is reported that three BLRGs, namely 3C~111, 3C~120, and 3C~390.3, showed blue-shifted Fe XXV/XXVI K-shell absorption lines in the {\it Suzaku} X-ray spectrum \citep{Tombesi10}. This feature is interpreted as highly ionized gas outflow whose velocity is of the order of $0.1c$. Since 3C~120 and 3C~111 are also detected by \fermi, these sources are ideal objects to study a physical link between accretion disk, ultra-fast outflow, and jet. Indeed, \citet{Tombesi11,Tombesi12} performed detailed study of 3C~111 by using X-ray and VLBA data \cite[see also][]{AH-WP19}.

Location of jet dissipation region is a long-standing matter of debate. To unveil this problem, multi-wavelength light curve and high spatial resolution VLBA images provide useful information. Here, we show {\it Fermi} Large Area Telescope monitoring result, together with 230~GHz Sub-Millimeter Array (SMA) and 43~GHz VLBA ones. Based on the light curves and broadband spectrum from radio to MeV/GeV $\gamma$ rays, we show that jet dissipation took place at sub-pc scale from the central BH and MeV/GeV $\gamma$ rays would be produced by synchrotron self Compton process, rather than inverse-Compton scattering of external photons from broad line region and dusty torus. Details are described in \citet{Tanaka15}.

\begin{figure*}[t]
\centering
\includegraphics[width=120mm]{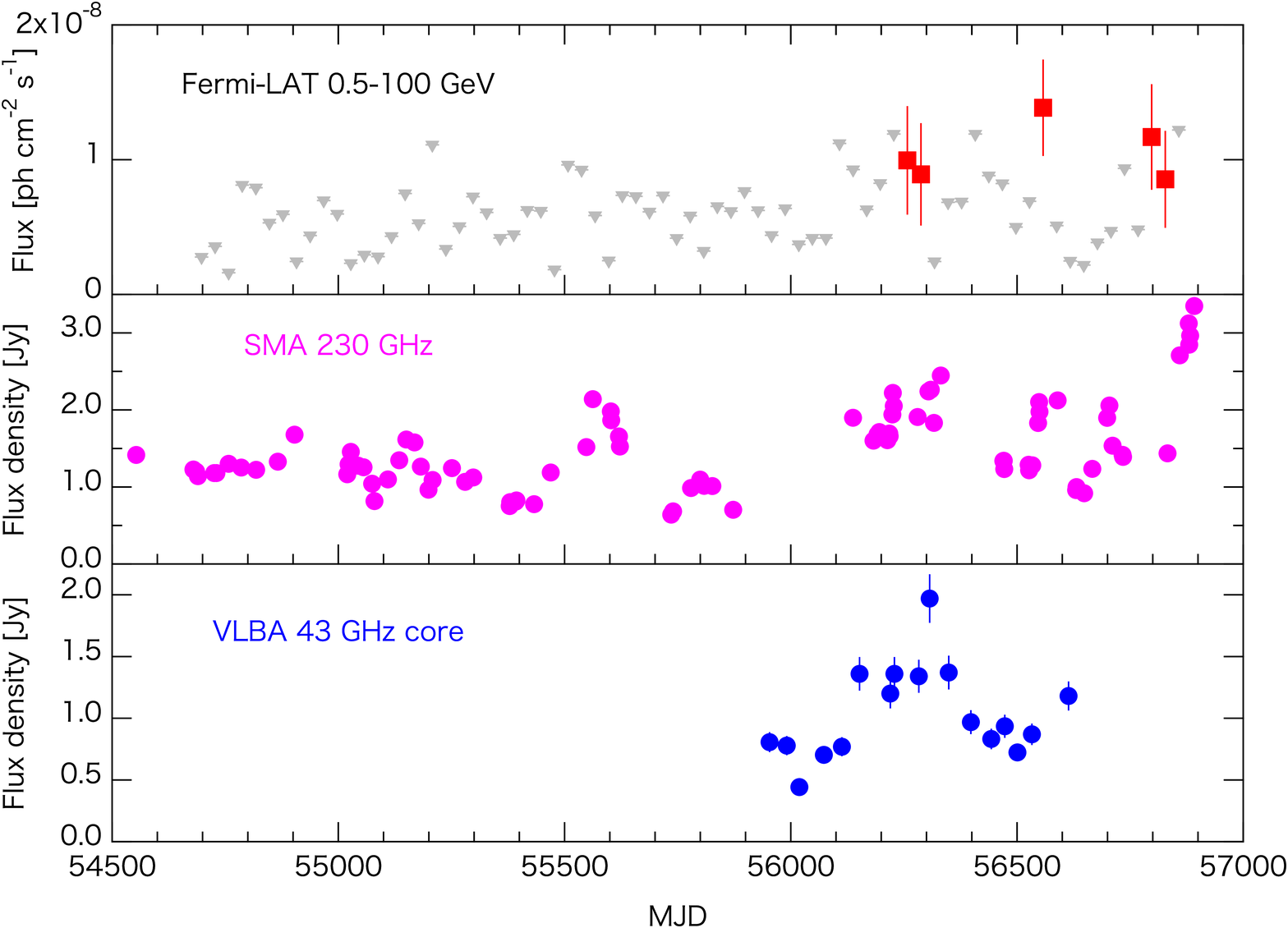}
\caption{\fermi\ (30-day bin), SMA, and VLBA light curves for \src\ since 2008 August. Gray triangles in the top panel show 90\% confidence level flux upper limits when ${\rm TS}<9$. Zoom-up light curve between MJD~56100 and 56500 are shown in Fig.~\ref{fig:lc2}.
Taken from \citet{Tanaka15}.} 
\label{fig:lc}
\end{figure*}

\begin{figure*}
\includegraphics[width=120mm]{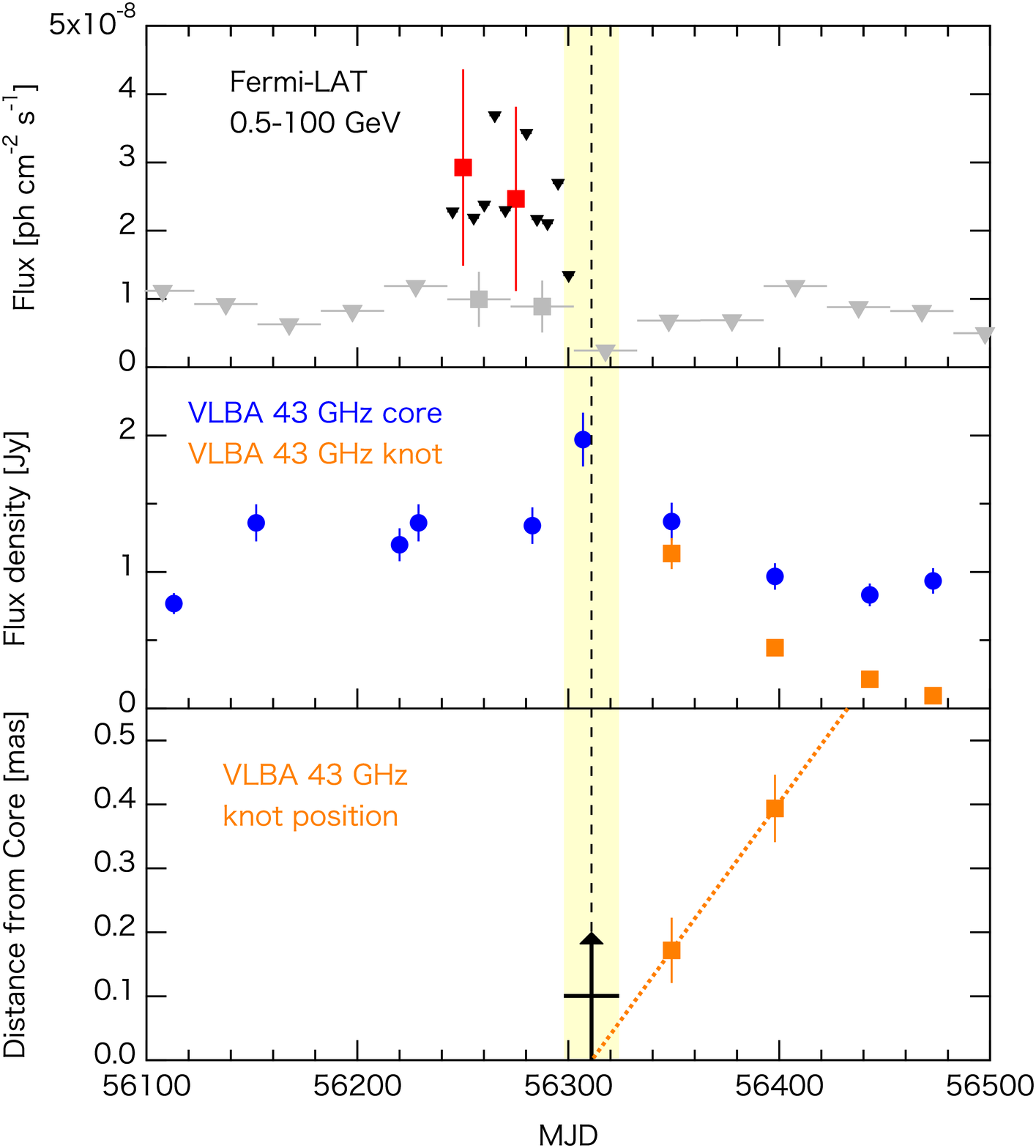}
\caption{Enlarged \fermi\ (5-day bin, top) and VLBA (mid) light curves between MJD~56100 and 56500. 
Black triangles in the top panel are 90\% confidence level flux upper limits when ${\rm TS}<4$. Gray squares and triangles are 30-day-bin flux and upper limits shown in the top panel of Fig.~\ref{fig:lc}.
The lower panel represents the relative position of the ejected knot from the central BH. Black arrow indicates the timing when the knot was ejected from the VLBA core.
Taken from \citet{Tanaka15}.}
\label{fig:lc2}
\end{figure*}

\section{Fermi-LAT and Multi-wavelength Observations}
Fig~\ref{fig:lc} shows the \fermi, 230~GHz SMA, and 43~GHz VLBA light curves. After the \fermi\ detection, VLBA light curve showed brightening. Closed-up light curves between MJD~56100 and 56500 are displayed in Fig.~\ref{fig:lc2}. MeV/GeV $\gamma$-ray emission is followed by core brightening and knot ejection.

\begin{figure*}[t]
\centering
\includegraphics[width=120mm]{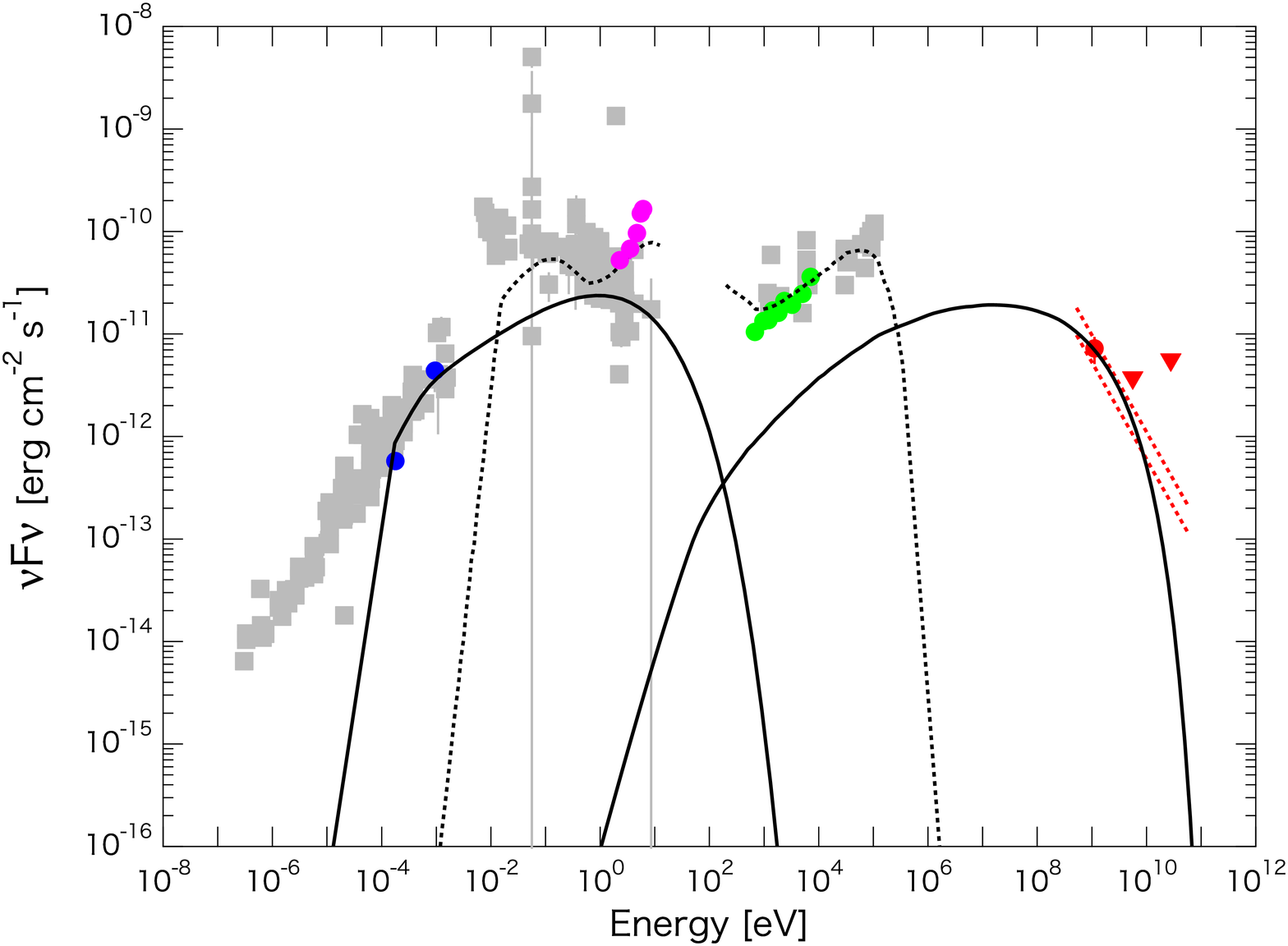}
\caption{Broadband spectrum of \src. Shown by gray are historical fluxes taken from NED database. 43 and 230~GHz fluxes measured on MJD~56283 and 56280, respectively, are shown by blue squares. Magenta and green are Swift/UVOT and XRT fluxes observed on MJD~56276. Red squares and triangles are \fermi\ flux and 95\% upper limits, derived from 60-day accumulated data from MJD~56242.7 to 56302.7. Solid line indicates one-zone synchrotron+SSC modeling, while dashed line are accretion-related model.
Taken from \citet{Tanaka15}.} 
\label{fig:sed}
\end{figure*}

Fig~\ref{fig:sed} shows a broadband spectrum from radio to GeV band. Swift/UVOT and XRT data points obtained during the GeV flaring state are also plotted together. Radio, sub-mm, and MeV/GeV fluxes are nicely fitted by one-zone synchrotron-self-Compton modeling (see Section~\ref{sec:dis}), while optical-UV and X-ray fluxes are obviously above the jet component and are reasonably represented by accretion-related disk model by \citet{Korat98}.

\section{Discussion}
\label{sec:dis}
We first assume that the MeV/GeV $\gamma$-ray enhancement and radio knot ejection event are physically connected. Then, our observation (the $\gamma$-ray detection before the core brightening and subsequent knot ejection) indicates that $\gamma$-ray emission region is located inside the VLBA core. From long-term X-ray and VLBA monitoring over six years, \citep{Chatterjee09} derived the distance from the central BH to 43~GHz VLBA core as $\sim$0.5~pc. Then, we can infer the location of the $\gamma$-ray emission region from the time lag ($\sim60$ and $\sim35$ days) and viewing angle of $\sim20.5^{\circ}$ as $\sim 0.1$ and $\sim 0.3$ pc from the central BH.

MeV/GeV $\gamma$ rays are thought to be produced by inverse Compton scattering. There are three candidate photons sources; synchrotron photons, photons from broad line region, and hot dusty torus. Since the $\gamma$-ray emission region is located far beyond the broad line region of $R_{\rm BLR}=0.019-0.024$~pc, which is derived by reverberation mapping \citep{Pozo14}, we can safely neglect the contribution of BLR photons as targets. By using reasonable parameter values, we obtain $L_{\rm ERC}/L_{\rm SSC} \approx 0.1$, indicating that synchrotron self Compton is favored \citep[see][for details]{Tanaka15}.

To derive the physical quantities, we performed SED modeling. See \citet{Tanaka15} for the derived parameter values, but here we briefly summarize the important points. The ratio of comoving electron and magnetic field density is obtained as $u^{\prime}_{\rm e}/u^{\prime}_{\rm B} \sim 0.4$, suggesting almost equipartition. Total radiated power estimated from SED modeling is $L_{\rm rad} \sim 4.9 \times 10^{44}$ erg s$^{-1}$. Given the observational fact that $L_{\rm jet} \simeq L_{\rm rad}$ for AGN jets \citep{Nemmen12}, we obtain $L_{\rm jet} \approx 5 \times 10^{45}$ erg s$^{-1}$. On the other hand, total power of accreting plasma $L_{\rm acc}$ is estimated as $L_{\rm acc} \approx 10 L_{\rm disk}=2 \times 10^{45}$ erg s$^{-1}$. Hence, we obtain $L_{\rm jet} \simeq L_{\rm acc}$, meaning the jet launch is extremely efficient \citep[e.g.][]{Tanaka11, Saito13, Ghisellini14}.

Finally, we mention a future perspective. Next-generation X-ray satellite {\it Astro-H} \citep{Takahashi12} is scheduled to be launched in 2015/2016. Ultra-high-resolution spectroscopy by micro-calorimeter SXS and simultaneous broadband spectral measurement by SXI, HXI, and SGD will provide new information about a physical link between accretion disk, ultra-fast outflow, and jet. In particular, highest sensitivity of SGD at 50--600~keV will enable us to detect the jet component in soft $\gamma$-ray band, as was detected by {\it Suzaku} from Centaurus~A core by \citep{Fukazawa11}. Hence, simultaneous \fermi\ and VLBA observations are also complementary and allow us to precisely derive physical parameters (such as electron distribution and magnetic field) at the emission region by SED modeling.

\section{Summary}
We presented six-year multi-wavelength light curve since the launch of \fermi\ in 2008 August\footnote{Similar \fermi\ analysis paper for \src\ by \citet{Sahakyan14} was posted on the arXiv during the review process of our original paper \citet{Tanaka15}.}. Under the assumption that MeV/GeV $\gamma$-ray flux increase is physically connected by the subsequent VLBA core brightening and knot ejection, we derived the $\gamma$-ray emission region is located at $\sim 0.1$ and $\sim 0.3$~pc from the central BH. We conclude that synchrotron-self-Compton process is preferred as a $\gamma$-ray emission mechanism. Future {\it Astro-H} observation will provide unique opportunity to study physical relation between accretion disk, ultra-fast outflow and relativistic jet.

\bigskip 
\begin{acknowledgments}
The \textit{Fermi}-LAT Collaboration acknowledges support for LAT development, operation and data analysis from NASA and DOE (United States), CEA/Irfu and IN2P3/CNRS (France), ASI and INFN (Italy), MEXT, KEK, and JAXA (Japan), and the K.A.~Wallenberg Foundation, the Swedish Research Council and the National Space Board (Sweden). Science analysis support in the operations phase from INAF (Italy) and CNES (France) is also gratefully acknowledged.

This study makes use of 43~GHz VLBA data from the VLBA-BU Blazar Monitoring Program (VLBA-BU-BLAZAR;
http://www.bu.edu/blazars/VLBAproject.html), funded by NASA through the Fermi Guest Investigator Program. The VLBA is an instrument of the National Radio Astronomy Observatory. The National Radio Astronomy Observatory is a facility of the National Science Foundation operated by Associated Universities, Inc. 
\end{acknowledgments}

\bigskip 


\end{document}